\documentclass{article}

\usepackage{graphicx}%
\usepackage{multirow}%
\usepackage{amsmath,amssymb,amsfonts}%
\usepackage{amsthm}%
\usepackage{mathrsfs}%
\usepackage[title]{appendix}%
\usepackage{xcolor}%
\usepackage{textcomp}%
\usepackage{manyfoot}%
\usepackage{booktabs}%
\usepackage{algorithm}%
\usepackage{algorithmicx}%
\usepackage{algpseudocode}%
\usepackage{listings}%
\usepackage{algorithm, algpseudocode}
\usepackage{array}
\usepackage[thinlines]{easytable}
\usepackage{tabularx}
\newcommand\ML[1]{\llap{#1\quad}}
\usepackage{caption}
\usepackage{subcaption}
\usepackage{setspace}
\usepackage{comment}
\usepackage{authblk}
\usepackage{hyperref}
 \hypersetup{
     colorlinks=true,
     linkcolor=blue,
     filecolor=blue,
     citecolor = black,      
     urlcolor=black,
     }
\providecommand{\keywords}[1]
{
  \small	
  \textbf{\textit{Keywords---}} #1
}
\title{Development Of Raspberry Pi-based Processing Unit for UV Photon-Counting Detectors}

\author[1,2]{Bharat Chandra P.}

\author[1]{Binukumar G.}

\author[1]{Shubham Janakiram}

\author[1]{Mahesh Babu}

\author[1]{Shubhangi Jain}

\author[1]{Richa Rai}

\author[1]{Rekhesh Mohan}

\author[1]{Margarita Safonova}

\author[1]{Jayant Murthy}

\affil[1]{ Indian Institute of Astrophysics, Bangalore, India}

\affil[2]{Department of Applied Optics and Photonics, University of Calcutta, Kolkata, India}
\date{}

\begin{document}
\maketitle
\begin{abstract}
In ultraviolet (UV) astronomical observations, photons from the sources are very few compared to the visible or infrared (IR) wavelength ranges. Detectors operating in the UV usually employ a photon-counting mode of operation. These detectors usually have an image intensifier sensitive to UV photons and a readout mechanism that employs photon counting. The development of readouts for these detectors is resource-intensive and expensive. In this paper, we describe the development of a low-cost UV photon-counting detector processing unit that employs a Raspberry Pi with its in built readout to perform the photon-counting operation. Our system can operate in both $3\times3$ and $5\times5$ window modes at 30 frames per sec (fps), where $5\times5$ window mode also enables the provision of detection of double events. The system can be built quickly from readily available custom-off-the-shelf (COTS) components and is thus used in inexpensive CubeSats or small satellite missions. This low-cost solution promises to broaden access to UV observations, advancing research possibilities in space-based astronomy. 
\end{abstract}


\section{Introduction}
\noindent 
Photon-counting detectors are widely used in UV astronomical observations, where the photons are relatively few but high energy. The most common detector system is a microchannel plate (MCP) based image intensifier with a UV-sensitive photocathode tailored to the wavelength of interest. Electrons from the MCP are directed to a readout, of which there are many different types. A review of UV detectors for space astronomy has been given by Joseph\cite{1995ExA.....6...97J}. These detector systems usually consist of a micro-channel-plate (MCP)-based image intensifier with a UV-sensitive photocathode and a detector (such as, e.g. CMOS (complementary metal-oxide semiconductor), delay-line, cross-strip anode, etc.) with the readout electronics. In the photon-counting mode of operation, the system scans the detector elements for photon events and determines the centroids of those events. Photon-counting detectors have the advantage of no read noise and low background counts, because MCP provides a huge gain for incoming photons during the avalanche multiplication, and the threshold set for detection eliminates all other background noises. UV space missions such as FUSE \cite{2000SPIE.4013..334S}, GALEX \cite{2005ApJ...619L...1M}, IUE \cite{1993UNPSA...4..173W}, ASTRO-1 \cite{1993Sci...259..327D} and UVIT \cite{2012SPIE.8443E..1NK} have employed photon-counting detectors for imaging and spectroscopy. GALEX used a 2D anode delay line detector readout placed behind the MCP for the field-programmable gate array (FPGA)-based processing. IUE used an image intensifier with a television camera with associated readout electronics to perform photon counting \cite{tv_cam}. Hopkins Ultraviolet Telescope (HUT) and Ultraviolet Imaging Telescope of the ASTRO-1 mission used MCP with a scanning diode array directly interfaced and processed with the onboard flight computer. FUSE had a Z-stack MCP, a double delay-line detector, and associated electronics to perform the photon counting. The UVIT detector uses a phosphorus screen to convert the electrons from the MCP into visible photons. These photons are imaged by a CMOS sensor with FPGA-based readout electronics to perform centroiding of the events. Detector development has typically been costly in terms of money and time, and we have been developing low-cost detectors suitable for use in CubeSat-class missions. Our previous effort used a near-ultraviolet (NUV) intensified detector from Photek, Inc., for which we developed a Spartan 6 FPGA readout \cite{2017JAI.....650002A}.

In most of the designs, the readout and centroiding are implemented using complex logic circuitry or FPGA. Among them, FPGA has the flexibility of reprogramming by changing the bit stream, which is not possible in the case of hard-implemented logic circuitry. The problem with these two approaches in implementing a photon-counting detector is the complexity of development, long development and testing time, limited modification flexibility, and large costs associated with all that. To overcome these problems, we propose an innovative photon-counting detector processing unit based on an embedded microprocessor platform. For this, we have used Raspberry Pi (RPi) cam V2 as the camera and a Raspberry Pi Zero 2W\footnote{\tt{https://www.raspberrypi.org/}} as an embedded microprocessor platform. Here, we take advantage of the inbuilt readout implemented in the RPi Graphics Processing Unit (GPU) to handle the camera readout. This embedded Linux platform-based design provides the flexibility of easy development, a short development cycle, less design complexity for the team, and easy code portability. The RPi Zero 2W has an inbuilt camera readout integrated into its GPU, which handles the readout operation and the processing is done by the CPU. This paper describes the development and testing of such a photon-counting detector using a Photek\footnote{\tt{https://www.photek.com/}} image intensifier, Raspberry Pi Zero 2W (SBC), and the RPi cam V2 camera to perform photon counting and centroid calculations. We have also implemented single-event recovery and system reset in case of a malfunction in the detector system. We believe that this inexpensive design can be very advantageous to use in small satellites or in CubeSat UV missions.

We have now made use of the rapidly improving capabilities of an embedded Linux microprocessor to develop a new readout based on a Raspberry Pi (RPi) cam V2 as the camera and a Raspberry Pi Zero 2W\footnote{\tt{https://www.raspberrypi.org/}} as an embedded microprocessor platform. This design provides the flexibility of easy development with a simpler design, short development cycles, and easy code portability. This paper describes the development and testing of such a photon-counting detector using a Photek\footnote{\tt{https://www.photek.com/}} image intensifier, embedded Linux Single-Board-Computer (SBC), and the CMOS camera to perform photon counting and centroid calculations. We have also implemented single-event recovery and system reset in case of a malfunction in the detector system. We believe this inexpensive and flexible design will be useful for small satellites and CubeSats.

\section{System Design}

\begin{figure}[h!]
    \centering
    \includegraphics[width=\linewidth]{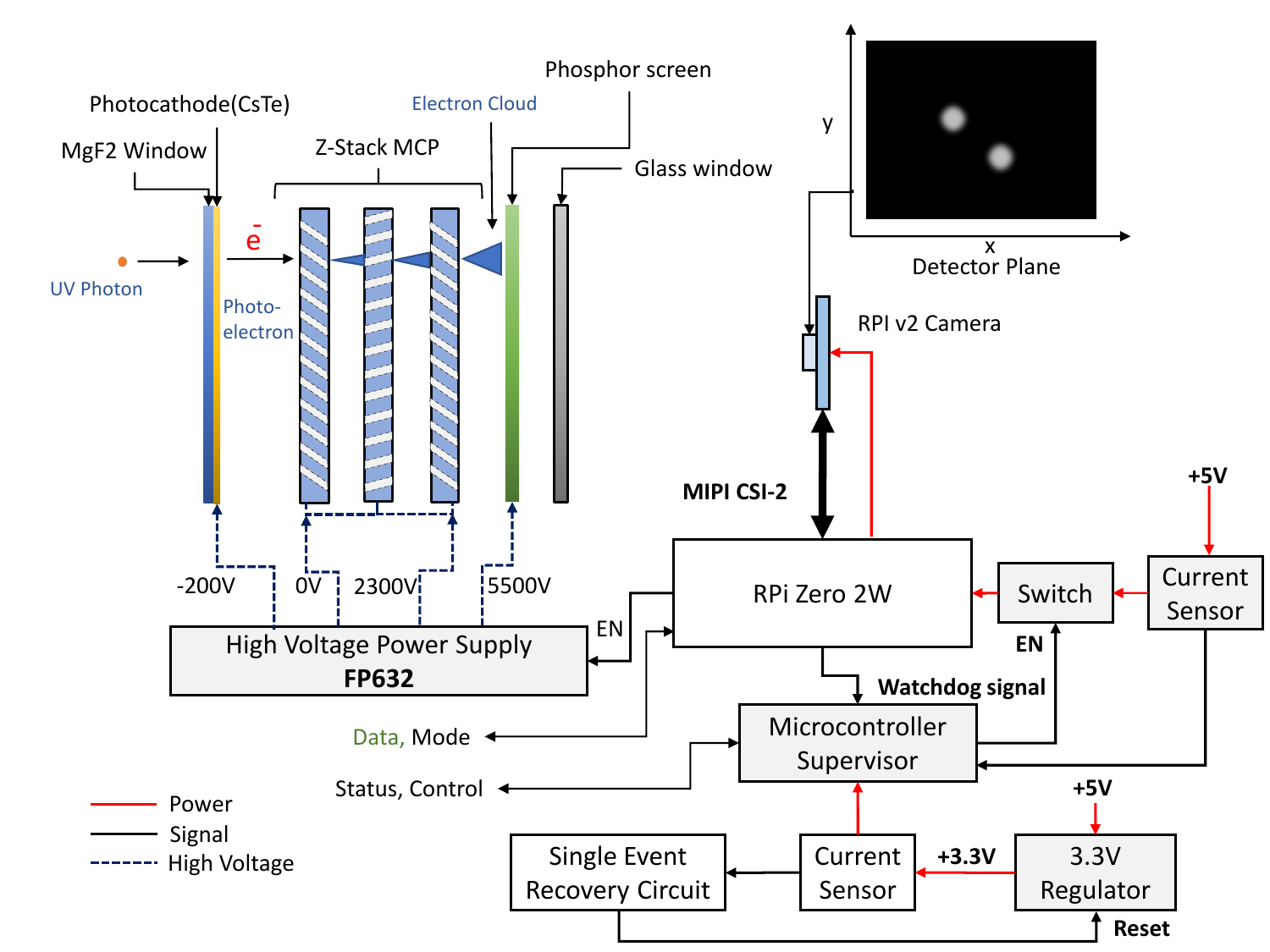}
    \caption{Block diagram of the photon-counting detector with high voltage power supply, single recovery circuit, Raspberry Pi and camera module.}
    \label{block_diagram}
\end{figure}
\begin{figure}[h!]
    \centering
    \includegraphics[width=\linewidth]{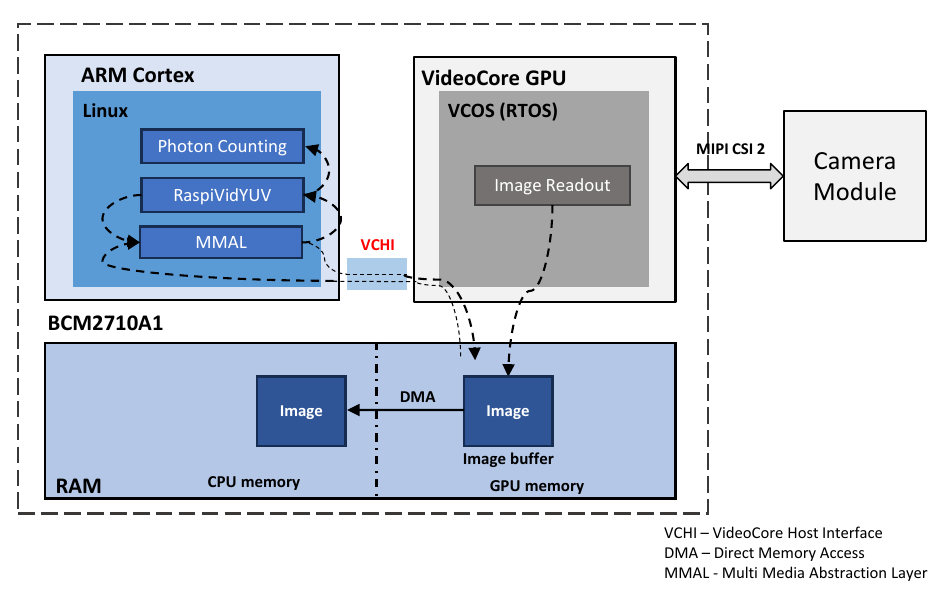}
    \caption{Block diagram of RPi camera readout and photon counting implementation using the inbuilt CPU and GPU.}
    \label{gpu_readout}
\end{figure}

\begin{figure}[h!]
    \centering
    \includegraphics[width=0.8\linewidth]{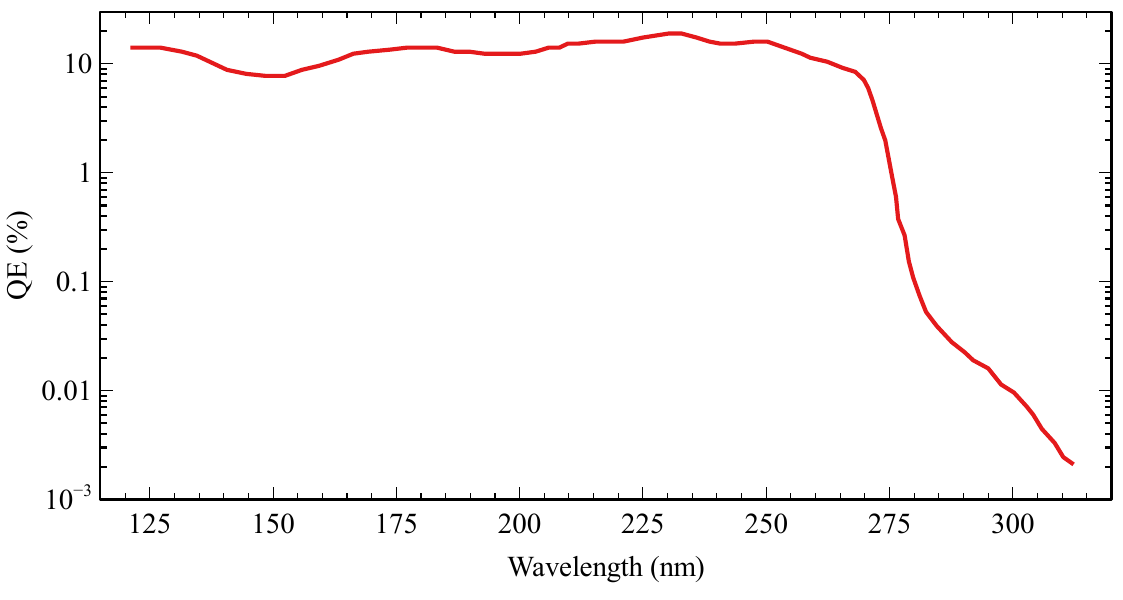}
    \caption{Quantum efficiency plot of the Photek CsTe MCP image intensifier.}
    \label{qe_plot}
\end{figure}

Our system requirement was to build a photon-counting readout for a planned NUV (1400 to 2700~\AA) mission \cite{2020SPIE11444E..76C} using a 40-mm MCP from Photek with a solar-blind caesium telluride (CsTe) photocathode( Fig.~\ref{qe_plot} shows the QE plot). Incoming UV photons falling on the photocathode generate photoelectrons, which are accelerated towards the MCP stack by a potential difference provided by the high-voltage supply. These electrons undergo avalanche multiplication in the MCP stack with a gain of $10^6 - 10^7$, and the resulting electron cloud falls on the phosphorus screen to generate visible light photons. We use the RPi V2 camera module to directly image the phosphor screen, and the RPi Zero 2W board to readout the image from the CMOS using inbuilt GPU readout and CPU to process the image. The microcontroller supervisor is the ATmega328p, which has been tested for a total ionizing dose (TID) of 56 kRad \cite{rpi_rad}. Fig.~\ref{block_diagram} shows the overall block diagram of the system.

\subsection{Camera and embedded Linux processing unit}

The RPi V2 camera operates was set in $1280\times720$-pixels video mode at 30 frames per second\footnote{The RPi Cam v2 supports up to 60 fps in 1280x720 video mode and up to 90 fps in 640x480 video mode.} with a fixed exposure time of 33 ms, which has been configured in the software. The processing unit is the RPi Zero 2w, running a minimal version of Raspbian. The RPi camera module was connected to the RPi's GPU through the MIPI-CSI-2 interface and controlled with the C++ RaspiVidYUV library\footnote{The RaspiVidYUV is a copy of the RaspiVid program which produces raw video in YUV output format.} \cite{raspividyuv}. Here, the readout operation is performed by the RPI's GPU, which runs an RTOS(Real Time Operating System) called VCOS. The video stream from the camera module is read out by the VideoCore GPU and written to RAM allocated to the GPU, and this frame is then transferred to the CPU side of the RAM using Direct memory access for processing. The entire operation is controlled using the VCHI(VideoCore Host Interface) and the underlying MMAL(Multi-Media Abstraction Layer) drivers of RasPiVidYUV. Fig.~\ref{gpu_readout} shows how the readout and processing operation is done in the Raspberry Pi\cite{Picamera}. We wrote image processing code in standard C/C++ to take the video from the RaspiVidYUV stream and perform the photon counting and centroiding operations, which were then output through the UART serial port. The centroiding and the UART transmission process were written as separate threads using the Pthreads Library \cite{pthread} to meet the real-time requirement. 
In the $3\times3$ window mode operation, we set the frame rate to 30 fps, exposure time to 33 ms, and operating mode to the continuous frame transfer mode. In the $5\times5$ window mode operation, we set the frame rate to 30 fps, exposure time to 33 ms, ISO to 800, and operating mode to also the continuous frame transfer mode.  

\begin{figure}[h!]
    \centering
    \begin{subfigure}[b]{0.48\textwidth}
    \includegraphics[width=\textwidth]{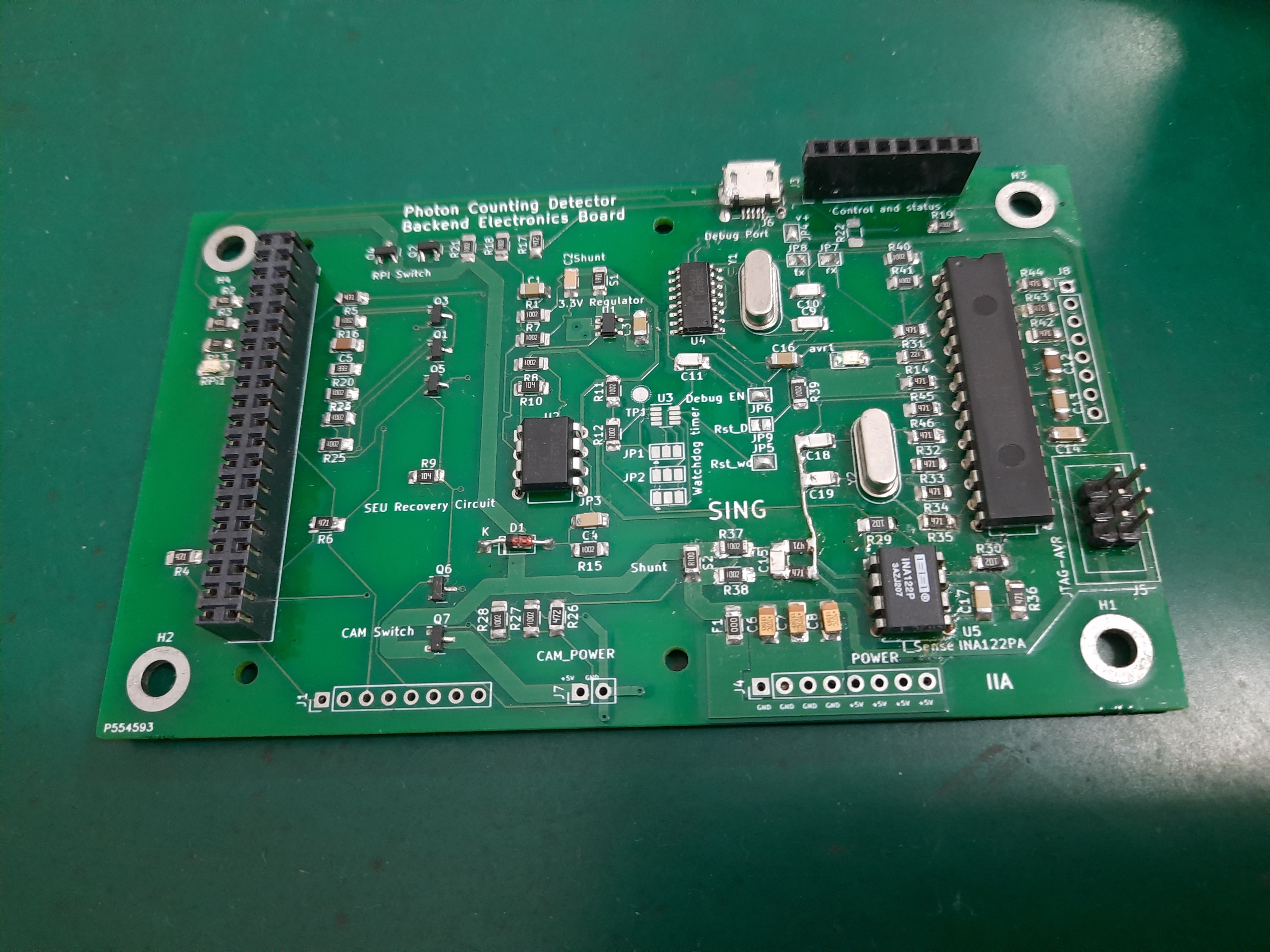}
    \caption{Recovery and protection circuit based on Atmel controller with current monitoring and watch dog feature.}
        \label{rc}
   \end{subfigure}
   \hfill
       \begin{subfigure}[b]{0.4\textwidth}
    \centering
    \includegraphics[width=\textwidth]{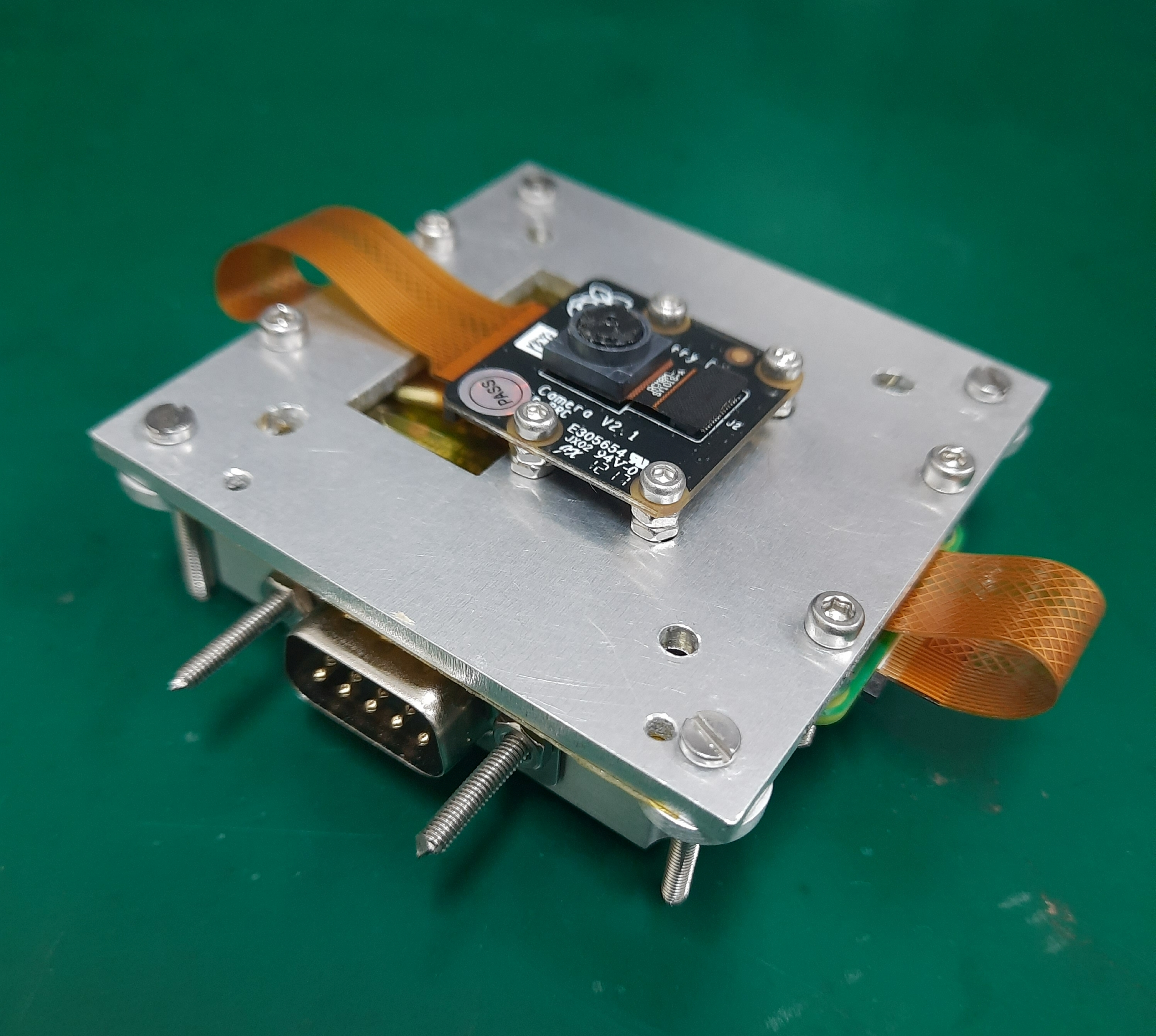}
   \caption{Enclosure for the the RPi camera V2 and RPi zero 2W with interface connector.}
   \label{cmos}
    \end{subfigure}
\caption{Images of the detector electronics board.}

\end{figure}

We also added a microcontroller board in the system to monitor the RPi operation and power cycle the RPi in case of any single event upset (SEU) in the system. The power supply for the microcontroller is independent and has an inbuilt single-event recovery circuit implemented with discrete electronics components \cite{rad_hard}. Fig.~\ref{rc} shows the recovery and protection circuit based on the Atmel controller, and Fig.~\ref{cmos} shows the RPi camera V2 and RPi zero 2W with an interface connector. Tables~\ref{cam_specs} and \ref{tab:rpi} show the camera and RPi Zero 2W specifications, respectively.
\begin{table}[h!]
\footnotesize
\begin{subtable}[c]{0.45\textwidth}
\begin{tabular}{ll}
\hline
Image Sensor & Sony IMX219 \\
Focal length & 3.04 mm \\
Focal ratio & F2.0 \\
Pixel size & 1.12 µm $\times$ 1.12 µm\\
Pixel Resolution & 3280 $\times$ 2464\\
Sensor format &  1/4 inch\\
\hline
\end{tabular}
\caption{RPi Cam V2 technical specifications.}
\label{cam_specs}
\end{subtable}  
\begin{subtable}[c]{0.5\textwidth}
\begin{tabular}{ll}
\hline
Chip Broadcom    & BCM2710A1                             \\
CPU              & ARM Cortex-A53 $\times$ 4 64-bit   \\                  
Operating system & Raspbian                            \\
CPU Clock        & 1 GHz                               \\
Memory           & 512 MB LPDDR2                \\
Interfaces       & $1 \times$Micro USB                 \\
                 & $1 \times$UART                      \\
I/O              & 40 GPIO Pins                        \\
                 &  CSI camera connector               \\
Onboard storage  & SD, MMC, SDIO card slot             \\
Power rating     & 250 mA (1.25 W)                     \\
Dimensions      & 66.0mm $\times$ 30.5mm $\times$ 5.0mm   \\
\hline
\end{tabular}
\caption{RPi Zero 2W board technical specifications.}
\label{tab:rpi}
\end{subtable}
\caption{Technical Specifications of RPi cam V2 (a), and Raspberry Pi Zero 2W (b).}
\end{table}   

\section{The Algorithm}
\noindent
The algorithm implemented here is based on the algorithm used onboard the UVIT \cite{2007PASP..119.1152H}. Algorithm~\ref{algo} shows the implementation in our detector. The system can operate in the $3\times3$ or $5\times5$ window mode, which can be selected during device initialisation. The $3\times3$ window mode is suitable for crowded fields where there is a large number of photon events. In contrast, the $5\times5$ window mode is suitable for better photometric accuracy.

\begin{algorithm}[H]
\footnotesize
\caption{Photon counting algorithm}
\label{code2}
\begin{algorithmic}[1]
\For {i=1 to $N_{y}-W$}
      \For {j=1 to $N_{x}-w$}
      \If {$P_{ij}$ $>$ $P_{Th}$ \textbf{and} $P_{ij}$ greater than the eight neighbouring pixels}
            \State Calculate sum of the pixels in the window as $P_{Energy}$
            \If {$P_{En}$  $>$ $P_{ETh}$ }
                 \State Calculate X and Y centroids for the event.
            \EndIf
     \EndIf
    \EndFor
\EndFor
\end{algorithmic}
\label{algo}
\end{algorithm}

The algorithm first scans for pixels with values above a threshold $P_{Th}$. If the pixel value is above the threshold, it checks that $P_{ij}$ is greater than all the neighbouring pixels in the window. If the second condition is also satisfied, we calculate the sum of pixels in the window and make sure that the sum is greater than the energy threshold $P_{ETh}$. If this final condition is also satisfied, we calculate the weighted centroid $x_{cen}$ and $y_{cen}$ using the following equations for a $3\times3$ window \cite{1999ASPC..164..392K},
\begin{equation}
    x_{cen} =\frac{C_3 - C_1} { \sum_{i=1}^{3} C_i} \,,
\end{equation}
where $C_1$, $C_2$ and $C_3$ are sums of respective columns in Table~\ref{3x3}, and
\begin{equation}
    y_{cen} =\frac{R_3- R_1} { \sum_{i=1}^{3}  R_i } \,,
\end{equation}
where $R_3$, $R_2$ and $R_1$ are sums of respective rows in Table~\ref{3x3}. 

For a $5\times5$ window, the centroids are calculated as 
\begin{equation}
    x_{cen} =\frac{2C_5+C_4-C_2-2C_{1}}{\sum_{i=1}^{5} C_i} \,,
\end{equation}
where $C_1$, $C_2$, $C_3$, $C_4$ and $C_5$ are sums of respective columns in Table~\ref{5x5}, and
\begin{equation}
     y_{cen} =\frac{2R_5+R_4-R_2-2R_1}{\sum_{i=1}^{5} R_i} \,,
\end{equation}
where 
$R_1$, 
$R_2$,
$R_3$,
$R_4$ and $R_5 $ are sums of respective rows in Table~\ref{5x5}.

\begin{table}[ht!]
\begin{subtable}[c]{0.45\textwidth}
\centering
\begin{TAB}(e,1cm,1cm){ccc}{ccc}   
     1 & 2 & 3 
\end{TAB}\\[-\normalbaselineskip]
\vspace{2mm}
\begin{TAB}(e,1cm,1cm){|c|c|c|}{|c|c|c|}   
\ML{1}  $a_{11}$ & $a_{12}$  & $a_{13}$    \\
\ML{2}  $a_{21}$ & $a_{22}$  & $a_{23}$    \\
\ML{3}  $a_{31}$ & $a_{32}$  & $a_{33}$    \\
\end{TAB}
\caption{$3\times3$-pixel window diagram.}
\label{3x3}
\end{subtable}
\begin{subtable}[c]{0.45\textwidth}
\centering
\begin{TAB}(e,1cm,1cm){ccccc}{ccccc}   
     1 & 2 & 3 & 4 & 5
\end{TAB}\\[-\normalbaselineskip]
\vspace{2mm}
\begin{TAB}(e,10mm,10mm){|c|c|c|c|c|}{|c|c|c|c|c|}  
\ML{1}  $a_{11}$ & $a_{12}$  & $a_{13}$ & $a_{14}$ & $a_{15}$  \\ 
\ML{2}  $a_{21}$ & $a_{22}$  & $a_{23}$ & $a_{24}$ & $a_{25}$  \\ 
\ML{3}  $a_{31}$ & $a_{32}$  & $a_{33}$ & $a_{34}$ & $a_{35}$ \\ 
\ML{4}  $a_{41}$ & $a_{42}$  & $a_{43}$ & $a_{44}$ & $a_{45}$ \\
\ML{5}  $a_{51}$ & $a_{52}$  & $a_{53}$ & $a_{54}$ & $a_{55}$ \\ 
\end{TAB}
\caption{$5\times5$-pixel window diagram.}
\label{5x5}
\end{subtable}
\caption{ Window diagram of $3\times3$ (a) and $5\times5$ (b) modes. }
\end{table}

\section{Integration and Testing}

\subsection{Threshold determination}

\begin{figure}[h!]
    \centering
    \begin{subfigure}[b]{0.49\textwidth}
    \includegraphics[width=\textwidth]{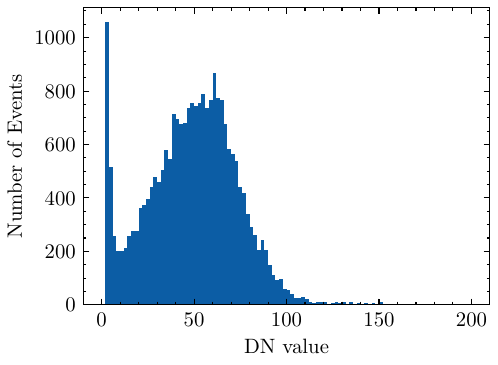}
    \caption{DN value histogram.}
    \label{1.1}
   \end{subfigure}
   \hfill
   \begin{subfigure}[b]{0.49\textwidth}
    \centering
    \includegraphics[width=\textwidth]{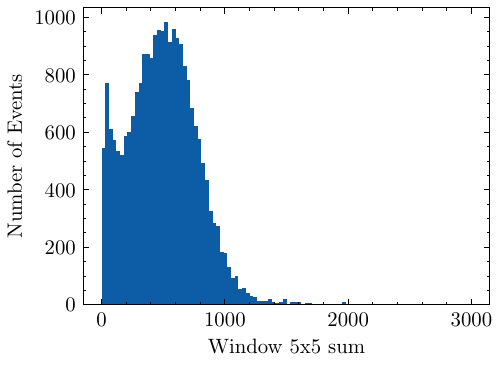}
    \caption{Sum of DN value histogram of $5\times5$ window mode.}
    \label{1.2}
    \end{subfigure}
    \begin{subfigure}[b]{0.49\textwidth}
    \centering
    \includegraphics[width=\textwidth]{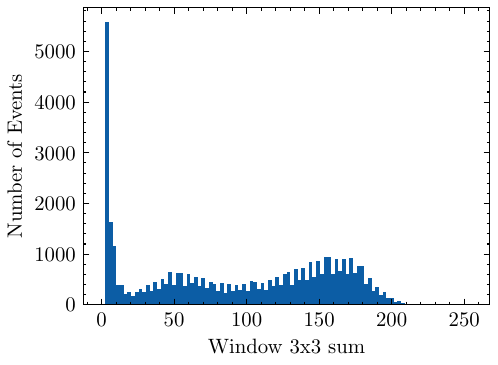}
    \caption{Sum of DN value histogram of $3\times3$ window mode.}
    \label{1.3}
    \end{subfigure}

\caption{Histogram plots of event height (a), energy distribution in $3\times3$ (b) and $5\times5$ (c) mode. }
\end{figure}

To determine the threshold, the MCP with RPi camera setup (Fig.~\ref{block_diagram}) was turned on in a dark room, with cathode voltage set to $-200$ V, MCP voltage to 2.6 kV, and phosphor screen voltage to 5.2 kV. Then we measured all events with peaks above 2 DN of the RPi V2 CMOS output, and calculate the sums of counts in both $3\times3$ and $5\times5$ windows. Fig.~\ref{1.1} shows the histogram distribution event peak, Fig.~\ref{1.2} shows the histogram of sum of events in $5\times5$ mode, and Fig.~\ref{1.3} shows the histogram of sum of events in $3\times3$ mode. We found that the mean background count of events was around 48.2 DN, so we used 50 as the pixel threshold. In the case of the energy threshold, we found that the mean sum for $5\times5$ mode was 495.3 DN, and 93.05 DN for the $3\times3$ mode. Therefore, we have selected 500 DN as the energy threshold for the $5\times5$ mode and 100 DN as the energy threshold for the $3\times3$ mode.

\subsection{Dark count estimation and double event detection}
\newcommand\darkCount{62.8}

CsTe photo-cathode MCP has a dark count of \darkCount \, photons/s (data provided by Photek). To verify this, we put a cover over the MCP to prevent any light from entering the photocathode windows. This was followed by measuring the counts every second as a function of time. We found that the average dark count value was 60.42 counts/s.

The $3\times3$ mode of operation is suitable for the observation of crowded fields and fields where less centroiding accuracy is needed. However, the $3\times3$ mode is not capable of detecting double events. The $5\times5$ mode has better centroiding accuracy, and a provision has been added to detect double events. This mode of operation sends the corner max and min pixel values along with centroids so that the user can use this data to flag double events based on the science case requirements \cite{2007PASP..119.1152H}.

\subsection{Air Force target imaging test}

\begin{figure}[h!]
    \centering
    \includegraphics[width=0.45\linewidth]{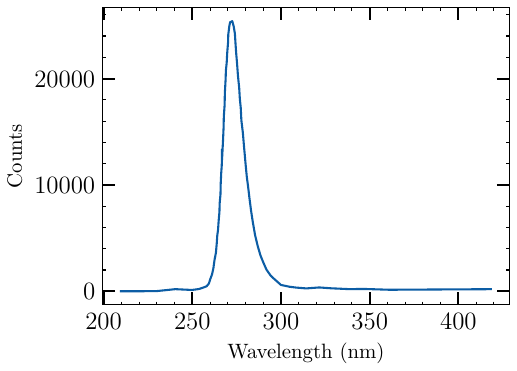}
    \caption{Spectrum of 275-nm UVC LED measured using MAYA2000 spectrometer.}
    \label{led}
\end{figure}

\begin{figure}[h!]
    \centering
    \includegraphics[width=\linewidth]{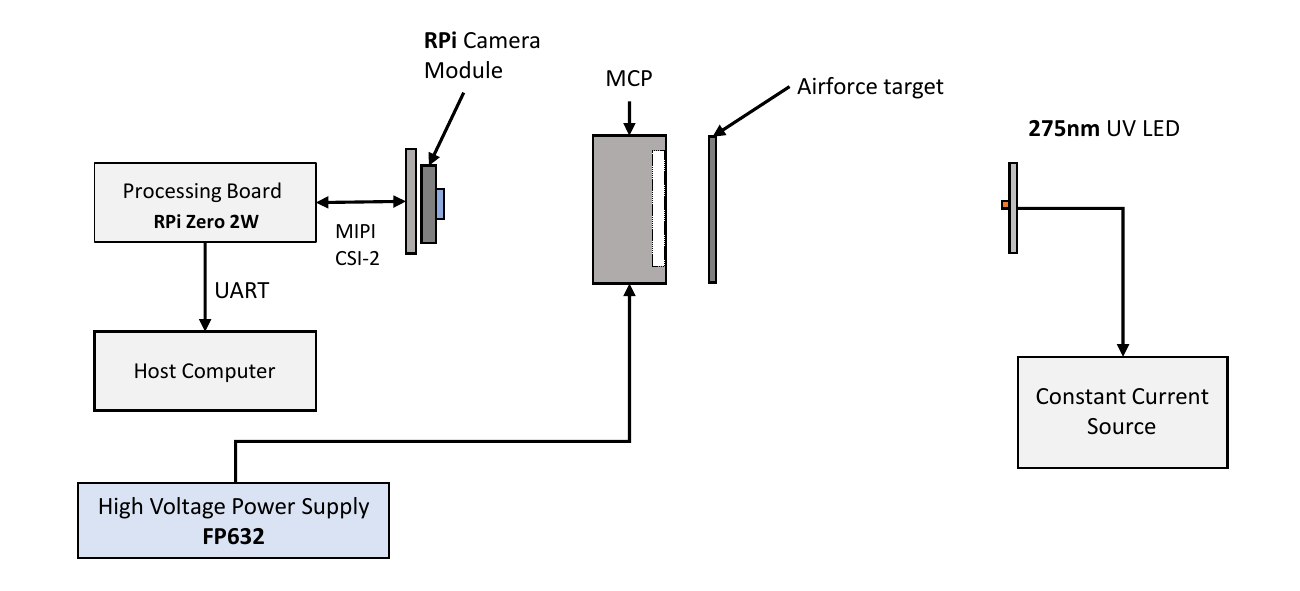}
    \caption{Block diagram of the Air Force target test.}
    \label{block_diagram_airforce}
\end{figure}

\begin{figure}[h!]
    \centering
    \begin{subfigure}[t]{0.54\textwidth}
    \includegraphics[width=\textwidth]{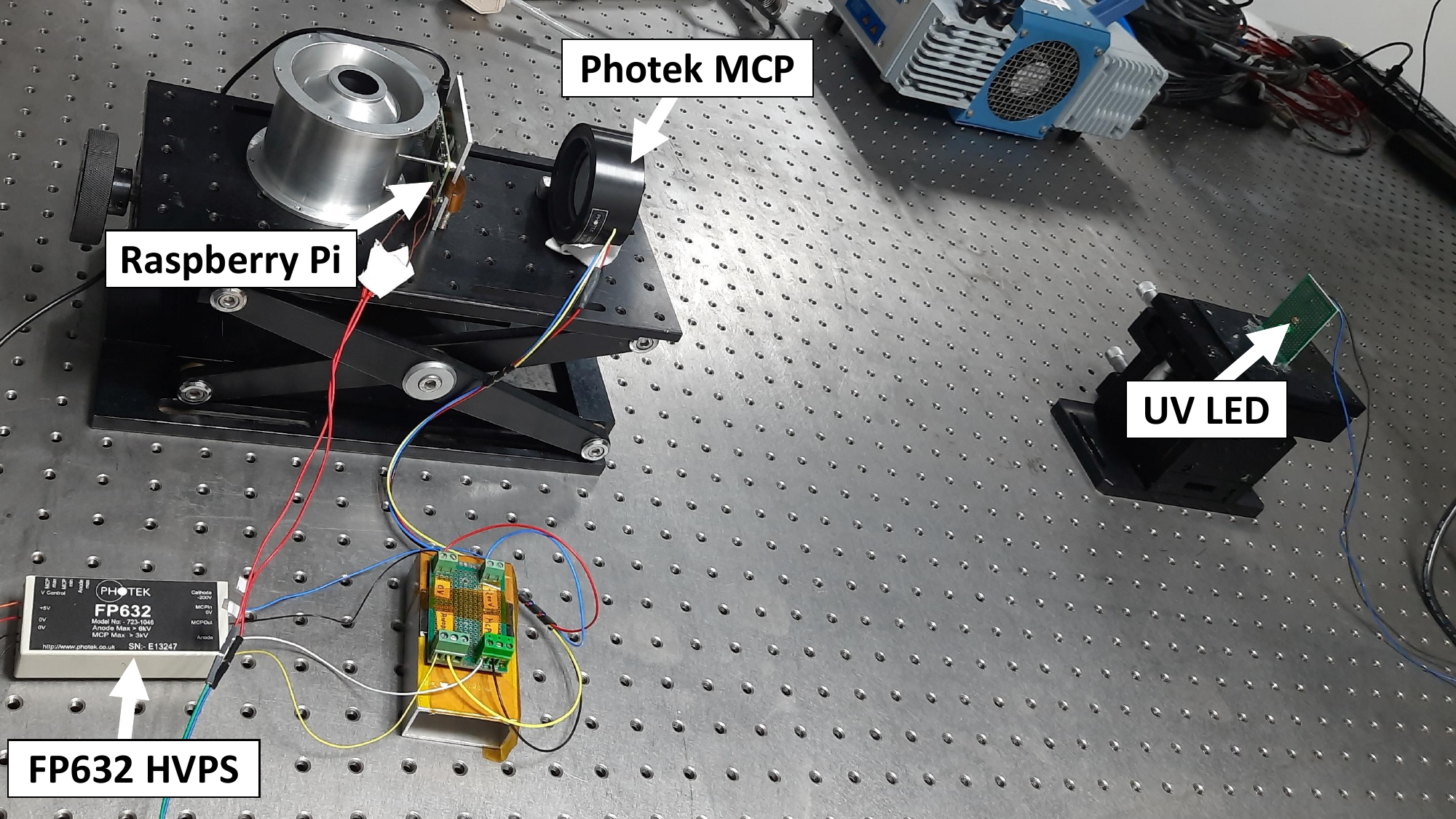}
    \caption{MCP imaging setup with UVC LED source and high-voltage power supply.}
   \end{subfigure}
   \hfill
\begin{subfigure}[t]{0.4\textwidth}
    \centering
    \includegraphics[width=\textwidth]{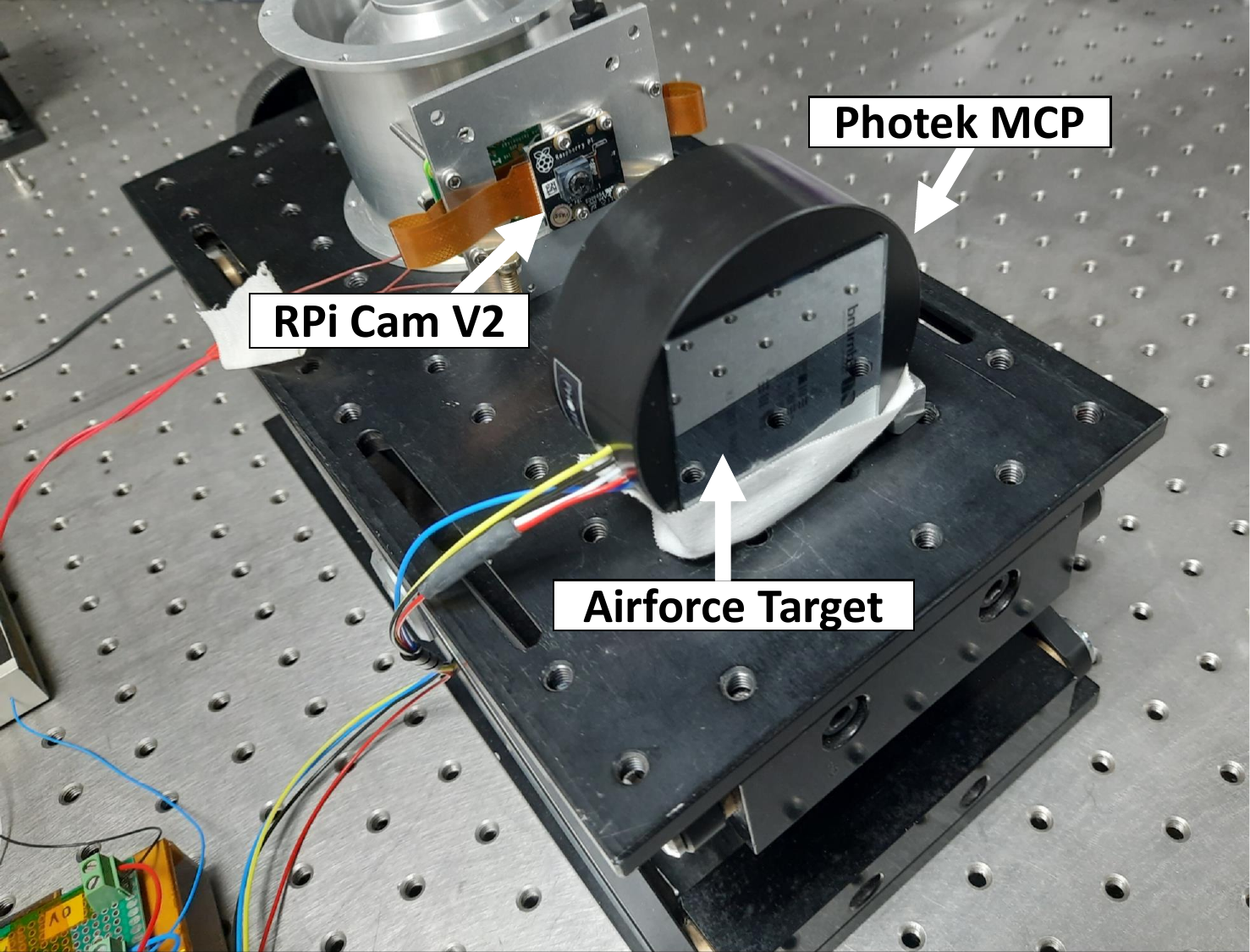}
   \caption{Detector window masked with Air Force target and RPi V2 camera imaging the phosphor screen.}
    \end{subfigure}

\caption{Detector imaging test setup.}
\label{set_up}
\end{figure}

\begin{figure}[h!]
    \centering
    \includegraphics[width=0.4\linewidth]{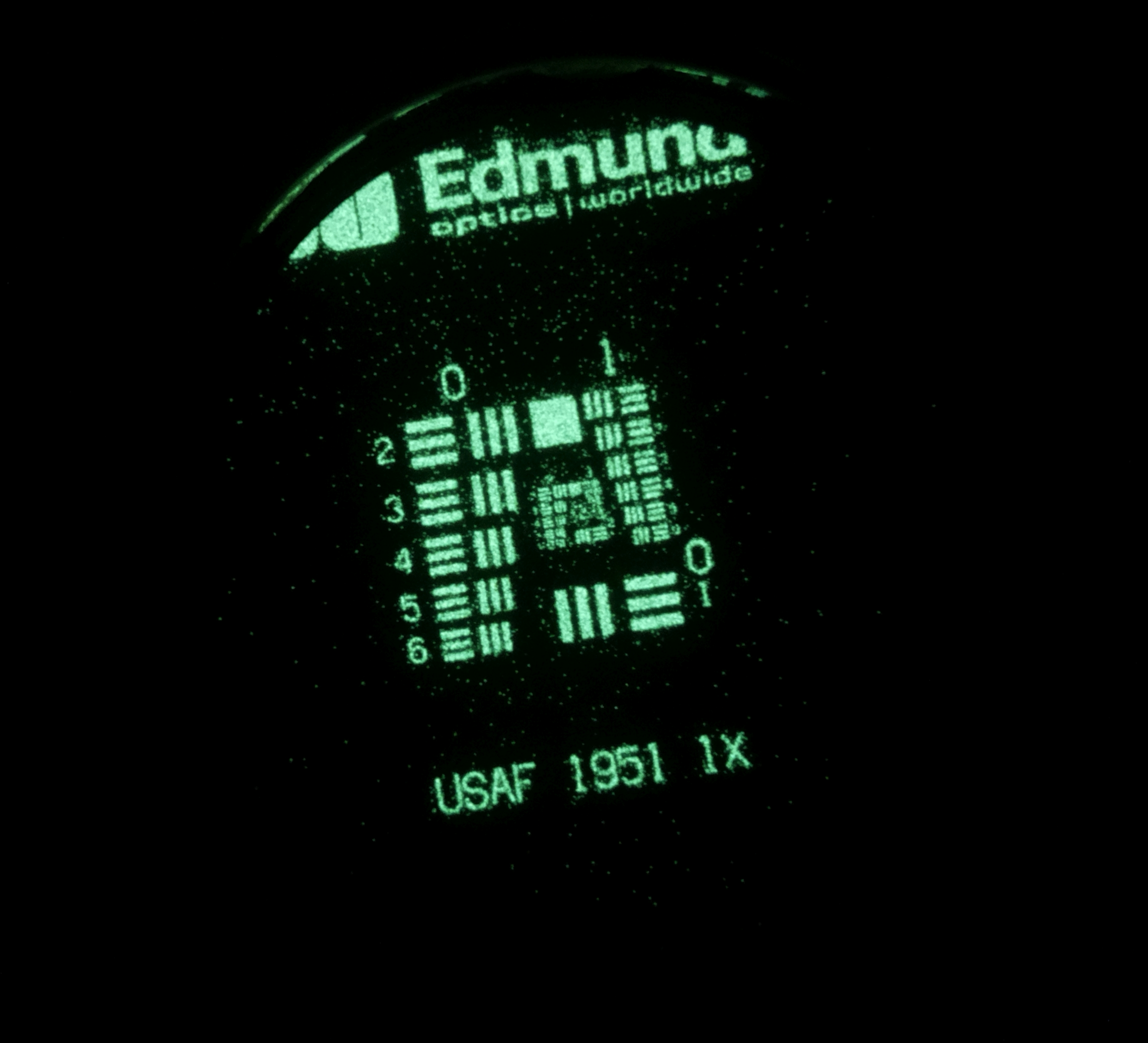}
    \caption{Direct image of the phosphor screen taken with the RPi camera when illuminated with 270-nm LED.}
    \label{actual}
\end{figure}

\begin{figure}[h!]
    \centering
    \begin{subfigure}[b]{0.49\textwidth}
    \includegraphics[width=\textwidth]{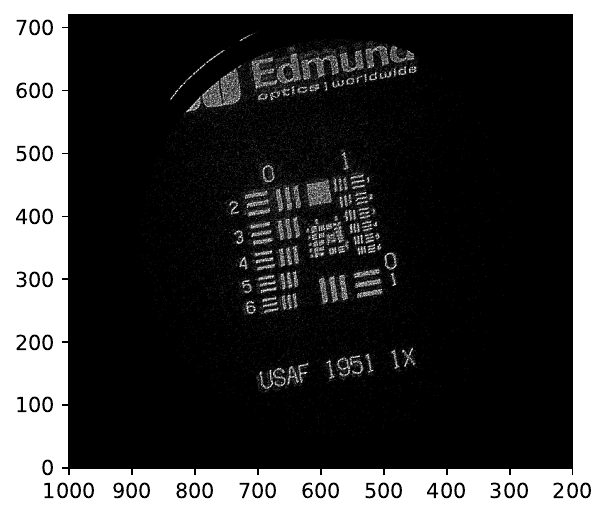}
    \caption{Image of Air Force target reproduced from the photon-counting data in $5\times5$ mode of operation.}
    \label{recon_A}
   \end{subfigure}
   \hfill
   \begin{subfigure}[b]{0.49\textwidth}
    \centering
    \includegraphics[width=\textwidth]{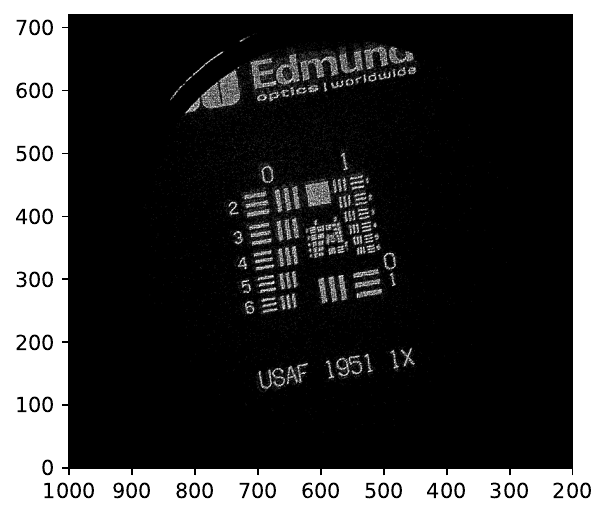}
    \caption{Image of Air Force target reproduced from the photon-counting data in $3\times3$ mode of operation.}
    \label{recon_B}
    \end{subfigure}
\caption{Images reconstructed from the centroids in $5\times5$ ({\it Left}), and $3\times3$ ({\it Right}) modes of operations for the air force target illuminated with the UV LED.}
\end{figure}

To perform the imaging test, the standard Air Force target was placed in front of the MCP window. Then the target was illuminated by a 275-nm UVC LED (3535UVC0.) controlled by an adjustable voltage source. Fig.~\ref{led} shows the spectrum of UV light emitted by the LED measured using MAYA2000
spectrometer\footnote{https://www.oceaninsight.com/products/spectrometers/high-sensitivity/Custom-Configured-maya2000-pro--series/} from Ocean Optics. Fig.~\ref{block_diagram_airforce} shows a block diagram of the setup used for the Air Force target imaging test and Figs.~\ref{set_up} show the lab setup used. Centroids, calculated by the RPi, were transmitted from the UART port at a 115200 baud rate, then captured on a PC using a Python script and stored in the PC memory. Later the image was reconstructed from the centroids on a $1280\times720$-pixel image grid. Figs.~\ref{recon_A} and \ref{recon_B} shows the reconstructed image from the centroids. Fig.~\ref{actual} shows a direct image of the phosphor screen taken with the RPi camera with long exposure.

\section{Assembled detector}

We have assembled a NUV detector using the concept described in the paper for our mission {\it SING}\cite{2020SPIE11444E..76C}, which is going to be a Near UV spectrograph operating from 140nm to 270nm.
A visual representation of the detector's cross-sectional view can be seen in Fig.~\ref{actual}(a), generated using CAD software. For housing the components, we utilized aluminium 6061-T6 to encapsulate the Photek MCP and the Raspberry Pi camera along with Raspberry Pi Zero 2W. The assembled detector is shown in Fig.~\ref{actual}(b), which was performed for the {\it SING} mission inside the M.G.K. Menon lab clean room facility at CREST.

\begin{figure}[h!]
    \begin{subfigure}[t]{0.48\textwidth}
    \centering
    \includegraphics[width=\linewidth]{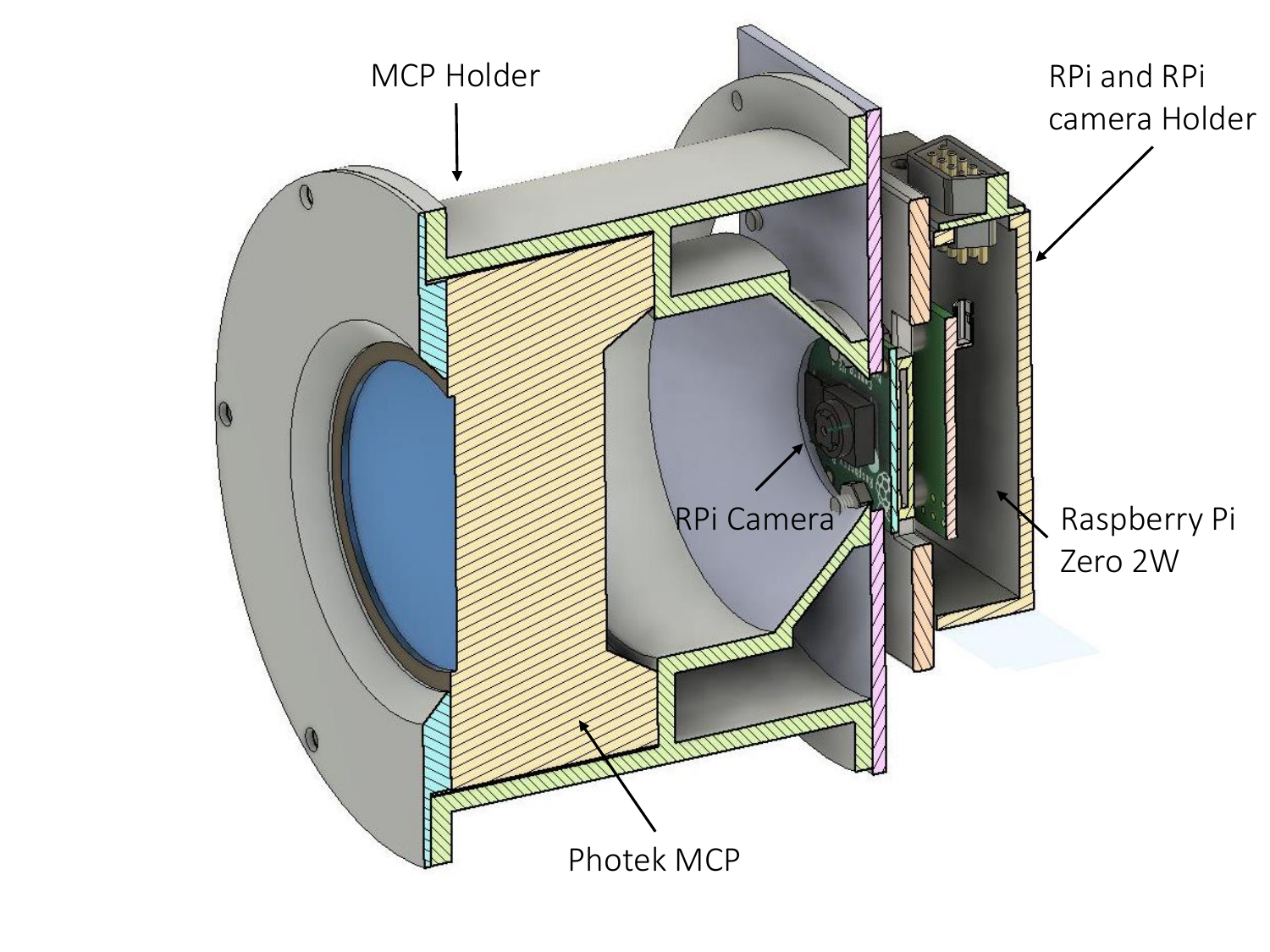}
    \caption{Section view of the detector in CAD }
   \end{subfigure}
   \hfill
       \begin{subfigure}[t]{0.48\textwidth}
    \centering
    \includegraphics[width=\linewidth]{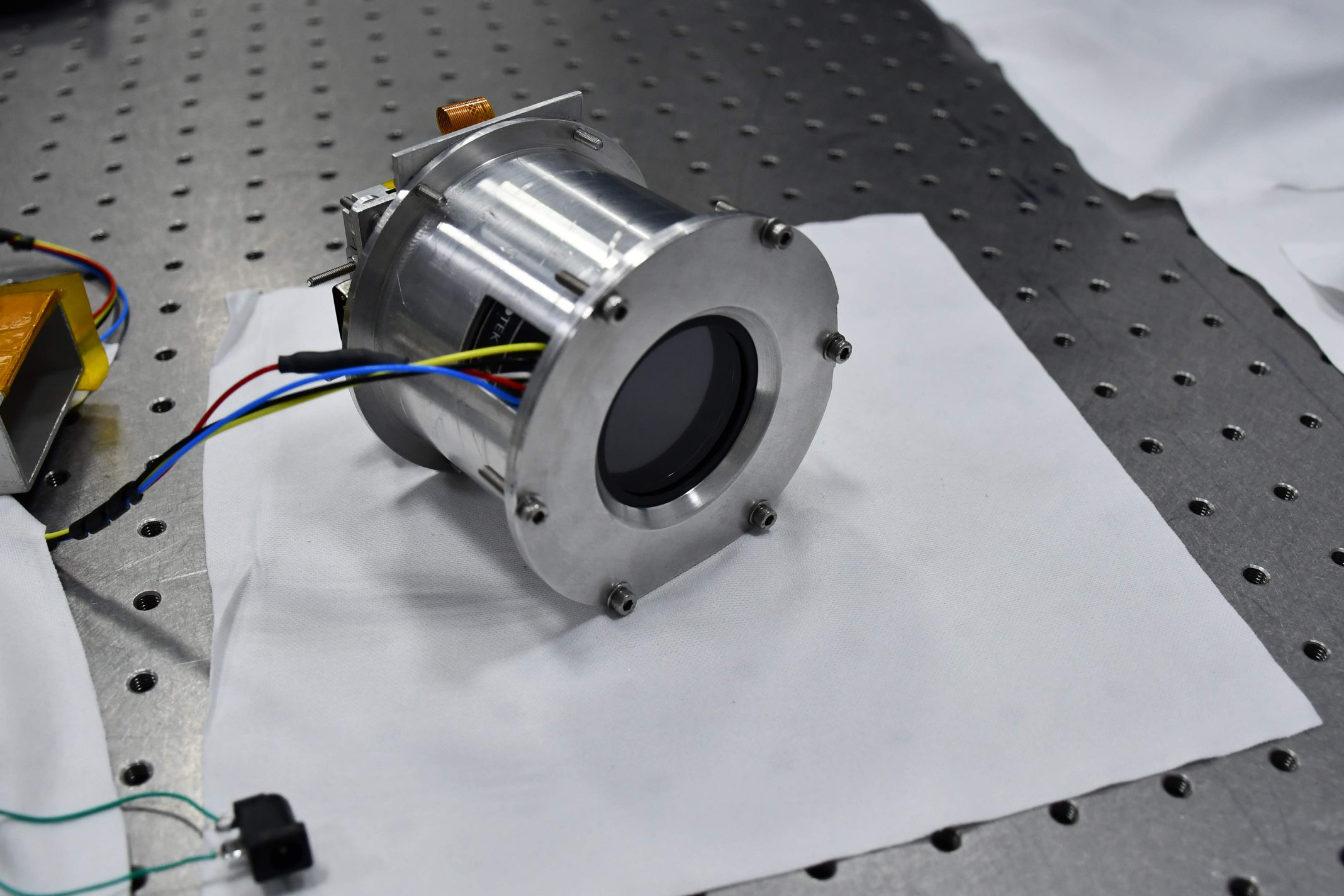}
    \caption{Assembled detector for the SING spectrograph payload.}
    \end{subfigure}

\caption{Assembled SING detector}
\label{Detector_assembly}
\end{figure}
\section{Conclusion and Future Work}

In this paper, we have presented the development of a photon-counting detector processing unit using Raspberry Pi, which can be used for UV astronomical observations. We are developing this Raspberry Pi-based electronics for our future small satellite UV missions. The detector electronics can perform photon counting in two modes of operation: $3\times3$ and $5\times5$. In $5\times5$ mode, we have implemented a mechanism for detecting double events. We have also demonstrated that photon-counting operations for UV detectors can be easily implemented on single-board computers such as Raspberry Pi. 

We will be using the same detector processing unit for the UV spectrograph {\it SING} that we are building at IIA \cite{2020SPIE11444E..76C}, which will operate in the wavelength range 140--270 nm. Once assembled, the entire detector system will undergo environmental and space qualification tests at M.~G.~K.~Menon laboratory (CREST) of IIA, Bangalore.

The electronics processing implementation using the Raspberry Pi is cheaper, faster and easily customisable as per the requirements.
The Raspberry Pi-based detector processing unit, along with its inbuilt readout, can be utilised in a wide range of other applications, such as soft X-ray synchrotron instrumentation, biological imaging, mass spectrometry, etc. 
  
\section*{Acknowledgments}

We wish to thank S. Sriram of the Indian Institute of Astrophysics (IIA) for his valuable suggestions and help. MS  acknowledges the financial support by the Department of Science and Technology (DST), Government of India, under the Women Scientist Scheme (PH), project reference number SR/WOS-A/PM-17/2019. We also thank all the staff at the M.G.K.Menon laboratory (CREST) for helping us with assembly in the cleanroom environment.

\section*{Code and data Availability}
\label{secA1}

The codes associated with this manuscript and information on its usage are available on \url{https://github.com/Bharatcs/Photon_counting_PI}. Any further data from the experiment will be available upon request.

\section*{Funding declaration} 

This research was made possible through financial support from the Indian Institute of Astrophysics (IIA) under the Department of Science and Technology (DST) in India.

\section*{Declaration of competing interest} 

The authors declare that they have no known competing financial interests or personal relationships that could have appeared to influence the work reported in this paper.

\bibliographystyle{ieeetr}
\bibliography{bibliography}

\end{document}